\documentclass[prb,twocolumn,showpacs,showkeys,superscriptaddress]{revtex4}
\usepackage{graphicx}
\usepackage{multirow}
\usepackage{longtable}

\begin{document}

\title{Random local strain effects in homovalent-substituted relaxor ferroelectrics: a~first-principles study of BaTi$_{0.74}$Zr$_{0.26}$O$_3$}
\author{C.~Laulh\'e}
\affiliation{Laboratoire des Mat\'eriaux et du G\'enie Physique (CNRS - Grenoble INP), MINATEC, 3 parvis Louis N\'eel, B.P. 257, F-38016 Grenoble cedex 01, France}
\affiliation{Synchrotron SOLEIL, L'Orme des Merisiers, Saint Aubin - B.P. 48, F-91192 Gif-sur-Yvette cedex, France}
\author{A.~Pasturel}
\affiliation{Laboratoire Science et Ing\'enierie des MAt\'eriaux et Proc\'ed\'es, 1130 rue de la Piscine, BP 75, F-38402 Saint Martin d'H\`eres cedex, France}
\author{F.~Hippert}
\author{J.~Kreisel}
\affiliation{Laboratoire des Mat\'eriaux et du G\'enie Physique (CNRS - Grenoble INP), MINATEC, 3 parvis Louis N\'eel, B.P. 257, F-38016 Grenoble cedex 01, France}

\date{\today}

\begin{abstract}
We present first-principles supercell calculations on BaTi$_{0.74}$Zr$_{0.26}$O$_3$, a prototype material for relaxors with a homovalent substitution. From a statistical analysis of relaxed structures, we give evidence for four types of Ti-atom polar displacements: along the $\left\langle 1 1 1 \right\rangle$, $\left\langle 1 1 0 \right\rangle$, or $\left\langle 1 0 0 \right\rangle$ directions of the cubic unit cell, or almost cancelled. The type of a Ti displacement is entirely determined by the Ti/Zr distribution in the adjacent unit cells. The underlying mechanism involves local strain effects that ensue from the difference in size between the Ti$^{4+}$ and Zr $^{4+}$ cations. These results shed light on the structural mechanisms that lead to disordered Ti displacements in BaTi$_{1-x}$Zr$_{x}$O$_3$ relaxors, and probably in other BaTiO$_3$-based relaxors with homovalent substitution.
\end{abstract}

\maketitle

The understanding of ferroelectric materials is a very active research area of constant interest over the past decades and with great relevance to both fundamental and application-related issues. \cite{rab07} Among ferroelectrics, relaxors form a fascinating class of materials which exhibit complex structural and dynamical behaviors on multiple time- and length- scales \cite{bok06}, with important technological applications such as capacitors and piezoelectric devices. \cite{par97}

Relaxors are characterized by a broad and frequency-dependent maximum of the dielectric permittivity as a function of temperature, in strong contrast with the sharp and frequency-independent anomalies observed in classic ferroelectrics. As the temperature decreases, most relaxors undergo no structural phase transition but develop a nanoscale structure that consists in polar atomic displacements with short-range correlations (so-called polar nanoregions). Chemical substitution was shown to be essential to obtain the relaxor properties, for it introduces spatial fluctuations of bonds and charges. \cite{bli00} The charge fluctuations, which occur in case of a heterovalent substitution ({\it e.g.} Mg$^{2+}$/Nb$^{5+}$ in the canonical relaxor PbMg$_{1/3}$Nb$_{2/3}$O$_3$), were long considered as the main driving force to determine the polar nanoregions. \cite{wes92} Nevertheless, the relaxor behavior is also observed in several lead-free BaTiO$_3$-based solid solutions with a homovalent substitution, as exemplified by BaTi$_{1-x}$Zr$_x$O$_3$ ($0.25~\le~x~\le~0.50$). \cite{sim04} Such systems are of great fundamental interest, as they offer an unique opportunity to isolate and study the role of lattice deformations in the appearance of the relaxor behavior.

In BaTi$_{1-x}$Zr$_x$O$_3$, the substituted Zr$^{4+}$ and Ti$^{4+}$ cations exhibit very different sizes ($r_{Ti^{4+}}$ = 0.605 {\AA}, $r_{Zr^{4+}}$ = 0.72 {\AA}). \cite{sha76} Such a size mismatch was shown to induce distortions of the oxygen atoms network in the classic-ferroelectric perovskite PbZr$_{1-x}$Ti$_x$O$_3$, as well as a reorientation of the Pb polar displacements depending on the Zr concentration. \cite{gri02,gri04} In BaTi$_{1-x}$Zr$_x$O$_3$ relaxors, the Zr substitution has to be thought as impeding the Ti polar displacement correlations that exist in the classic ferroelectric BaTiO$_3$, thus leading to the formation of polar nanoregions. However, the exact structural mechanisms involved have not yet been conclusively elucidated in BaTi$_{1-x}$Zr$_x$O$_3$, nor in other BaTiO$_3$-based relaxors. \cite{gli96,far00,shv06} Earlier experimental studies of the local structure in BaTi$_{1-x}$Zr$_x$O$_3$ showed that the O$_6$ octahedral cages around a Zr atom have a larger size ($\sim$~4.20~{\AA}) than those around a Ti atom ($\sim$ 4.05~{\AA}) \cite{lau06,lau09}, as expected from the different Ti$^{4+}$/Zr$^{4+}$ cation sizes. In order to know if a Zr atom further affects its surrounding structure, including the position of its Ti neighbors, one has to analyze the oxygen cage distortions and the Ti displacement changes as a function of the local Ti/Zr distribution. This information cannot be found from experiments since the atomic positions are averaged out on all the Zr and/or Ti octahedral sites, even when using local probes such as x-ray absorption spectroscopy (XAS) or pair distribution functions (PDFs). On the other hand, supercell first-principles calculations offer a promising (theoretical) route to study the local interplay between the Zr- and Ti-atoms. \cite{gri02,gri04}

\begin{figure}
\[\includegraphics[viewport=30 580 560 745,scale=0.45]{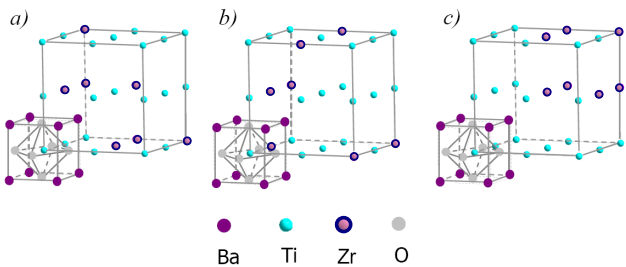}\]
\caption{(Color online) Arrangements of the 20 Ti- and 7 Zr-atoms in the BaTi$_{0.74}$Zr$_{0.26}$O$_3$ supercells used for our calculations. Only one of the 27 perovskite unit cells is fully represented to facilitate the visualization of the supercells.} \label{supercells}
\end{figure}

To model Ti/Zr substitution effects in the BaTi$_{1-x}$Zr$_x$O$_3$ relaxors, we used three 135-atom supercells representative of BaTi$_{0.74}$Zr$_{0.26}$O$_3$. The end-member compounds BaTiO$_3$ and BaZrO$_3$ were also considered as references to validate the calculation method. The initial atomic positions correspond to an assemblage of $3\times3\times3$ perovskite unit cells in their ideal cubic form (space group $Pm\bar3m$), with a lattice constant determined as the cube root of the experimental cell volume at 300 K (4.054, 4.006, and 4.192 {\AA} for BaTi$_{0.74}$Zr$_{0.26}$O$_3$, BaTiO$_3$, and BaZrO$_3$, respectively \cite{ver58,kwe93,mat91}). For BaTi$_{0.74}$Zr$_{0.26}$O$_3$, three prototype configurations of the 20 Ti- and 7 Zr-atoms were considered. In the first supercell (Fig. \ref{supercells}a) the average number of Zr neighbors of a Ti atom is 1.5 out of the 6 nearest octahedral B-sites, which is close to the value of 1.56 expected in the case of randomly distributed Zr atoms. The second (Fig. \ref{supercells}b) and third (Fig. \ref{supercells}c) supercells were built in such a way that the average number of Zr neighbors of a Ti atom is 1.7 (enhanced probability for Zr-Ti pairs of neighbors) and 1.2 (segregation of the Ti and Zr atoms), respectively. Various geometrical arrangements of the Zr atoms were considered: triangular arrangements parallel to the (111) planes (supercell ``a''), lines along the [100] directions (supercell ``b''), square arrangements parallel to the (100) planes (supercell ``c''), etc.

Total energies calculations were carried out using the density functional theory as implemented in the Vienna ab-initio simulation package. \cite{kre96a} Projected augmented plane waves \cite{kre99} (PAWs) with the Perdew-Wang exchange-correlation potential were adopted. The valence state of each element had been defined previously in the provided PAW potentials and the plane-wave cutoff was 400 eV. Numerical integrations in the Brillouin zone were performed by means of the Hermite-Gaussian method with N=1 and smearing parameter of $\sigma$=0.1 eV. Brillouin-zone sampling using a $4\times4\times4$ $k$-point grid was found necessary for differences of total energies of our supercells to converge within 10$^{-3}$ eV. Structural optimizations were carried out under the conditions that all residual forces should be smaller than 0.01 eV/{\AA}. The BaTi$_{0.74}$Zr$_{0.26}$O$_3$ supercell edges were constrained to form a cube with an edge-length of $3 \times 4.054 =$ 12.162~{\AA}, in order to match the crystallographic structure determined experimentally. \cite{ver58} For BaTiO$_3$ and BaZrO$_3$, only the supercell volume was fixed to the experimental value, the supercell-edge lengths and angles being free to vary.

In the relaxed supercells of BaTiO$_3$ and BaZrO$_3$, the atomic positions match within 1.5\% those determined by neutron diffraction at low temperatures. \cite{kwe93,akb05} We did not find any distortion with respect to the crystallographic structures, which is in agreement with the experimental studies of the local structure. \cite{com68,has95} This allows us to consider that the calculation method chosen is well suited for studying these two compounds and, by extension, their solid solution.

\begin{figure}
\[\includegraphics[viewport=60 400 535 710,scale=0.45]{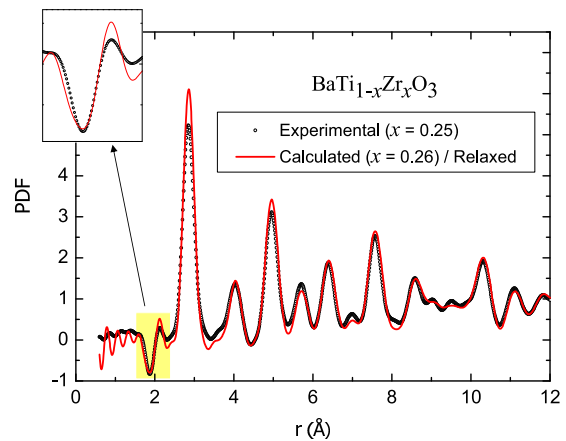}\]
\caption{(Color online) Experimental pair distribution function of BaTi$_{0.75}$Zr$_{0.25}$O$_3$ at 300 K (open circles) \cite{lau09}, compared with the PDF calculated from the three relaxed supercells of BaTi$_{0.74}$Zr$_{0.26}$O$_3$ (line).} \label{Compg(r)}
\end{figure}

Concerning BaTi$_{0.74}$Zr$_{0.26}$O$_3$, the energies of the relaxed structures are similar for the three supercells drawn in Fig. \ref{supercells}, within 0.31 eV (2.3 meV/atom). This energy difference is small enough to consider that each of the ``a'', ``b'', and ``c'' supercells potentially represents a small region of the real material. We find that the interatomic distance distributions calculated for the three BaTi$_{0.74}$Zr$_{0.26}$O$_3$ supercells are very similar, which suggests the existence of a single structural response to the Ti/Zr substitution, applicable for any local distribution of the Ti and Zr atoms. The PDF calculated from the summed interatomic distance distributions is compared in Fig. \ref{Compg(r)} with the experimentally determined neutron-PDF of BaTi$_{0.75}$Zr$_{0.25}$O$_3$ at 300 K (Ref. \onlinecite{lau09}). \cite{footnote} An excellent agreement is obtained between the two, especially in the $r$-range [1.5~-~2.5 {\AA}] that corresponds to the Ti-O and Zr-O distances within oxygen octahedra, in which the most important deviations with respect to the crystallographic structure were observed. \cite{lau09} Hence, the relaxed supercells capture the essential features of the local structure in the BaTi$_{0.74}$Zr$_{0.26}$O$_3$ relaxor, including the response of the Ti-O$_6$ and Zr-O$_6$ octahedra structures to the Ti/Zr substitution.

\begin{figure}
\[\includegraphics[viewport= 270 345 325 755,scale=0.45]{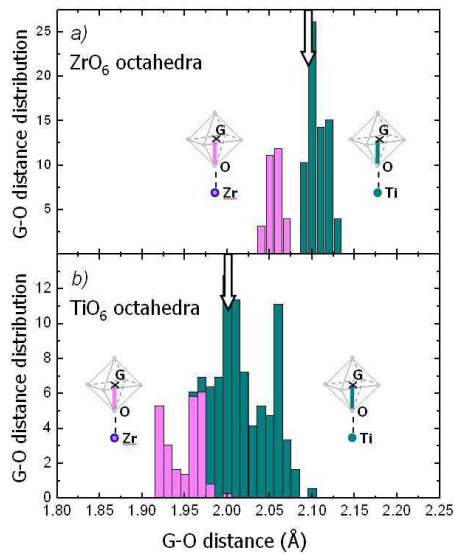}\]
\caption{(Color online) Distribution of the distance between the center of mass of oxygens (denoted G) and the oxygens calculated from the three relaxed supercells of BaTi$_{0.74}$Zr$_{0.26}$O$_3$: {\it (a)} for the Zr-O$_6$ octahedra, and {\it (b)} for the Ti-O$_6$ octahedra. The G-O segments that are directed toward Zr and Ti neighbors are represented in light magenta and dark green, respectively. The arrows mark the mean G-O distances calculated for the Zr-O$_6$ and Ti-O$_6$ octahedra.} \label{distriBO}
\end{figure}

We first determine the deformations undergone by the oxygen octahedra's network due to the Zr/Ti substitution. Defining G as the center of mass of an octahedron, the octahedral distortions can be classified into two categories: angular distortions, which result in a tilt of the G-O segments with respect to the Cartesian axes of the supercells, and radial distortions, which result in a change~of the G-O distances. The tilt angles of the G-O segments are found to be small, within the range [$0~-~1~^{\rm{o}}$]. They are not uniform in a given octahedron, and thus unambiguously correspond to angular distortions as opposed to global rotations of the octahedra. No relation is found between the tilt angles and the local chemical distribution. On the other hand, the G-O distances exhibit a marked dependence on both the~type (Ti-O$_6$ or Zr-O$_6$) of the considered octahedron and the local Zr/Ti distribution on its neighboring sites. From the calculated G-O distance distributions represented in Fig. \ref{distriBO}, one can indeed observe that the average G-O distance is larger in the Zr-O$_6$ octahedra (2.092 {\AA}) than in the Ti-O$_6$ octahedra (2.005 {\AA}). This result, which is expected from the different Ti$^{4+}$/Zr$^{4+}$ cation sizes, is in quantitative agreement with the previous experimental studies of the local structure. \cite{lau06,lau09} Besides, the large width of the peaks observed on the G-O distance distributions suggests that the shape of the octahedral cages fluctuates from one site to the other. In order to understand whether the local Zr/Ti distribution determines or partly determines this disorder, we considered separately the G-O distances that form a G-O-Ti chain (in dark green in~Fig. \ref{distriBO}), and the G-O distances that form a G-O-Zr chain (in light magenta in Fig. \ref{distriBO}). We find that~the~shortest G-O distances are observed when the neighboring octahedral site is occupied by a Zr atom, while the longest ones occur when the neighboring octahedral site is occupied by a Ti atom. The octahedra are thus compressed in the direction of Zr neighbors and expanded in the direction of Ti neighbors, which provides a microscopic demonstration of chemical pressure to which we will come back later.

\begin{figure}
\[\includegraphics[viewport=265 335 325 628,scale=0.6]{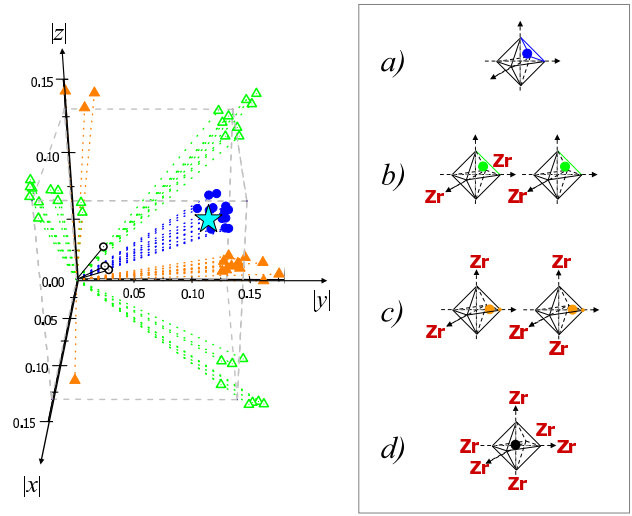}\]
\caption{(Color online) Ti displacement vectors calculated in the three relaxed supercells of BaTi$_{0.74}$Zr$_{0.26}$O$_3$, represented with the absolute value of their coordinates. Four types of displacement with distinct symmetries are observed: {\it a)} toward a face of the surrounding octahedron (full circles), {\it b)} toward its edge (open triangles), {\it c)} toward its corner (full triangles), and {\it d)} almost canceled (open circles). For comparison, the Ti displacement vector calculated in the classical ferroelectric BaTiO$_3$ is represented by a star. The symmetry of a Ti displacement is entirely determined by the distribution of Zr and Ti atoms in the adjacent unit cells, as illustrated in the inset. For the sake of clarity, only the Zr neighbors are represented for a given TiO$_6$ unit.} \label{DeplTi}
\end{figure}

It can be expected that the aforementioned distortions of the oxygen atom's network modify the bonding of the cations in their octahedral and cubo-octahedral cages, and hence their polar displacement. In order to investigate this point we calculated the cation displacement vectors away from the center of mass of their surrounding oxygen cages. The Ba and Zr atom displacements have a low average amplitude (0.06 {\AA} for Ba, 0.03 {\AA} for Zr), and their orientation is ill-defined. These characteristics indicate that the Ba and Zr atoms form mostly ionic bonds in the relaxor BaTi$_{0.74}$Zr$_{0.26}$O$_3$, as it is the case in the end-members BaTiO$_3$ and BaZrO$_3$. \cite{ber66,coh92} The average displacement amplitude calculated for the ferroelectrically active Ti atom in BaTi$_{0.74}$Zr$_{0.26}$O$_3$ is large (0.17 {\AA}) and nearly as high as the one calculated in the classical ferroelectric BaTiO$_3$ (0.18 {\AA}), which is in agreement with X-ray absorption near-edge structure studies of BaTi$_{1-x}$Zr$_{x}$O$_3$ at the Ti $K$-edge. \cite{lau07} However, the Ti displacement directions calculated for BaTi$_{0.74}$Zr$_{0.26}$O$_3$ and BaTiO$_3$ differ strongly. Figure \ref{DeplTi} gives a representation of the calculated Ti displacement vectors, the x-, y-, and z-axes being parallel to the Cartesian axes of the supercell ({\it i.e.}, the octahedra's axes before structural relaxation). The Ti displacements in BaTiO$_3$, which are all the same in amplitude and direction, are represented by the star. They have equal components along the x-, y-, and z-directions, and are thus directed along a $\left\langle 1 1 1 \right\rangle$ direction, toward a~face of their oxygen octahedron. In the relaxed supercell of BaTi$_{0.74}$Zr$_{0.26}$O$_3$, such Ti displacements toward an octahedron's face are still observed (full circles), but other types of displacement also appear: along $\left\langle 1 1 0 \right\rangle$ toward an octahedron's edge (open triangles), along $\left\langle 1 0 0 \right\rangle$ toward an octahedron's corner (full triangles), and even nearly cancelled (open circles). Starting from a displacement toward an octahedron's face, the three other cases can be obtained by successively cancelling one displacement component along the x-, y-, or z-axes. By analyzing the Ti atom environments in the relaxed supercells of BaTi$_{0.74}$Zr$_{0.26}$O$_3$, we establish that a Ti displacement component is cancelled as soon as a neighboring Zr atom lies on the corresponding octahedron's axis. This leads us to the important conclusion that in BaTi$_{0.74}$Zr$_{0.26}$O$_3$, the symmetry of a Ti-atom displacement is entirely determined by the distribution of the Ti and Zr atoms in the adjacent unit cells. Considering the octahedral distortions caused by the Zr/Ti substitution, we can also conclude that a Ti atom is no longer displaced in the direction in which its octahedral cage is compressed. Hence, the strain induced by the Zr/Ti substitution leads to a reorganization of the covalent bonds between the Ti atoms and their oxygen neighbors in BaTi$_{0.74}$Zr$_{0.26}$O$_3$, which results in a reorientation of the Ti polar displacements. 

It is instructive to set our results into the general context of the effect of strain in ferroelectric perovskites. More than thirty years ago, Samara {\it et al.} established that hydrostatic pressure (isotropic strain) reduces and even annihilates, at critical pressure, the correlations of polar cation displacements and thus ferroelectricity. \cite{sam00} Individual polar displacements are also sensitive to the pressure parameter. This is illustrated by XAS measurements on BaTiO$_3$, which showed that the Ti polar displacements are reduced to zero due to the compression of the octahedra along their three axes. \cite{iti06} As a second illustration of the effect of strain, let us remind that compressive biaxial strain in perovskite thin films can greatly enhance the remanent polarization in the direction perpendicular to the biaxial strain. \cite{cho04} Finally, we can cite the case of PbTiO$_3$, where the Ti-O$_6$ octahedra are elongated along one of their axes, due to the large displacement of Pb atoms. In this case, the local Ti displacements are directed in the direction of elongation, i.e. along a $\left\langle 1 0 0 \right\rangle$ direction rather than the preferred $\left\langle 1 1 1 \right\rangle$ ones. \cite{coh92,rav98} It is interesting to note that the previously reported strain effects concern uniform distortions throughout the oxygen atoms' network, in opposition to our here reported random local strain effects induced by chemical disorder. Nevertheless, chemical pressure in BaTi$_{0.74}$Zr$_{0.26}$O$_3$ acts on the individual polar displacements with similar rules to those observed for uniform lattice distortions.

Our results provide insight into the structural mechanisms that lead to disordered Ti displacements in BaTi$_{1-x}$Zr$_{x}$O$_3$ relaxors, and probably in other BaTiO$_3$-based relaxors with a homovalent substitution. We show that the chemical substitution imposes a random distribution of the Ti displacement symmetries among the Ti-O$_6$ octahedra, which impedes a perfect alignment of all the Ti displacements as it exists in the classic ferroelectric BaTiO$_3$ at low temperatures. \cite{com68} Nevertheless, the local distribution of the substituted cations does not completely determine the Ti displacements: a displacement toward an octahedron's face can be directed in eight possible directions, four possible directions for a displacement toward an octahedron's edge, and two possible directions for a displacement toward an octahedron's corner. This degree of freedom allows partial correlations of the polar displacements, and thus the formation of polar nanoregions. We believe that the reported behavior of the Ti polar displacements and its simple link to the local chemical order will give valuable constraints on both phenomenological and structural models which will aim at describing polar correlations in BaTiO$_3$-based relaxors.

\bibliographystyle{apsrev}

\begin{thebibliography}{99}
\bibitem{footnote} {The comparison between the PDFs measured at 300 K and calculated at 0 K can be justified on the basis of XAS experiments at the Ti and Zr $K$-edges, which exclude a significant temperature dependence of the local structure in BaTi$_{1-x}$Zr$_{x}$O$_3$ relaxors (Refs. \onlinecite{lau06,lau09}). Note that the calculated PDF was weighted by the suitable neutron scattering lengths, and convoluated with a Gaussian function as described in Ref. \onlinecite{lau09} in order to take into account the effect of thermal motions at 300~K.}
\end{thebibliography}

\end{document}